\documentclass[%
reprint,
superscriptaddress,
amsmath,amssymb,
aps,prl
]{revtex4-2}

\usepackage{graphicx}
\usepackage{dcolumn}
\usepackage{bm}
\usepackage{amsmath,amssymb}
\usepackage{multirow}
\usepackage[usenames,dvipsnames]{xcolor}
\usepackage{overpic}
\usepackage{pict2e}
\usepackage{mathtools}
\usepackage{ulem}
\usepackage{hyperref}

\newcommand{\pd}[2]{\frac{\partial #1}{\partial #2}}

\newcommand{\dd}[2]{\frac{\mathrm{d}\, #1}{\mathrm{d}\, #2}}
\newcommand{\pdd}[2]{\frac{\partial^2 #1}{\partial {#2}^2}}

\newcommand{\bracket}[1]{\left< #1 \right>}

\newcommand{\scrm}[1]{\mbox{\scriptsize #1}}
\newcommand{\fl}[1]{\tilde{#1}}
\newcommand{\rd}{\,\mathrm{d}}
\newcommand{\kB}{k_\mathrm{B}}
\newcommand{\ff}[1]{#1_\mathrm{ff}}
\newcommand{\fw}[1]{#1_\mathrm{fw}}

\definecolor{ggreen}{rgb}{0.0, 0.5, 0.0}

\newcommand{\green}[1]{\textcolor{ggreen}{#1}}

\newcommand{\ilm}{Univ Lyon, Univ Claude Bernard Lyon 1, CNRS, Institut Lumi\`ere Mati\`ere, F-69622, VILLEURBANNE, France}
\newcommand{\iuf}{Institut Universitaire de France (IUF), 1 rue Descartes, 75005 Paris, France}
\newcommand{\osaka}{Department of Mechanical Engineering, Osaka University, 2-1 Yamadaoka, Suita, Osaka 565-0871, Japan}
\newcommand{\tokyo}{Water Frontier Research Center (WaTUS), 
Research Institute for Science \& Technology,
Tokyo University of Science, 1-3 Kagurazaka, Shinjuku-ku, Tokyo 162-8601, Japan}
\newcommand{\osakac}{Department of Mechanical Engineering, 
Osaka City University, 3-3-138 Sugimoto, Sumiyoshi, Osaka, Osaka 558-8585, Japan}

\begin{document}
\title{Theoretical framework for the atomistic modeling of frequency-dependent liquid-solid friction}

\author{Haruki Oga}
\email{haruki@nnfm.mech.eng.osaka-u.ac.jp}
\affiliation{\osaka}
\author{Takeshi Omori}
\email{omori@osaka-cu.ac.jp}
\affiliation{\osakac}
\author{Cecilia Herrero}
\affiliation{\ilm}
\author{Samy Merabia}
\affiliation{\ilm}
\author{Laurent Joly}
\affiliation{\ilm}
\affiliation{\iuf}
\author{Yasutaka Yamaguchi}
\email{yamaguchi@mech.eng.osaka-u.ac.jp}
\affiliation{\osaka}
\affiliation{\tokyo}
\date{\today}

\begin{abstract}
Nanofluidics shows great promise for energy conversion and desalination applications. 
The performance of nanofluidic devices is controlled by liquid-solid friction, quantified by the Navier friction coefficient (FC).
Despite decades of research, there is no well-established generic framework to determine the frequency dependent Navier FC from atomistic simulations. Here, we have derived analytical expressions to connect the Navier FC to the random force autocorrelation on the confining wall, from the observation that the random force autocorrelation can be related to the hydrodynamic boundary condition, where the Navier FC appears. 
The analytical framework is generic in the sense that it explicitly includes the system size dependence and also the frequency dependence of the FC, which enabled us to address (i) the long-standing plateau issue in the evaluation of the FC and (ii) the non-Markovian behavior of liquid-solid friction of a Lennard-Jones liquid and of water on various walls and at various temperatures, including the supercooled regime. 
This new framework opens the way to explore the frequency dependent FC for a wide range of complex liquids.
\end{abstract}

\maketitle

\textit{Introduction}\textemdash Nanofluidics is the discipline that describes fluid motion in nano-confinement, whose unique behavior in the mass and ionic transport should be a key ingredient in 
future technologies for fluid filtration and energy harvesting \cite{Sparreboom2009,Bocquet2010,Daiguji2010,Faucher2019,Kavokine2021}. The recent advent of new materials and fabrication techniques to create fine fluid conduits \cite{Bocquet2020} has even increased the importance to explore nanofluidic transport. In nanofluidic systems, surface effects play a critical role because of the large surface-to-volume ratio. In particular,  liquid-solid slip can boost the performance of nanofluidic devices \cite{Joly2004,Ajdari2006,Ren2008,Radha2016}. 

Liquid-solid slip was first foreseen by Navier \cite{Navier1823}, who proposed that the slip velocity $u_{\scrm{slip}}$ is proportional to the shear stress on the wall $\tau$ as
\begin{equation}\label{eq:navier}
	\tau = \lambda\, u_{\scrm{slip}}
\end{equation}
with $\lambda$ 
the Navier friction coefficient (FC). 
The Navier boundary condition, Eq.~\eqref{eq:navier}, has been tested by many authors in the past decades 
as reviewed in the articles \cite{Neto2005,Bocquet2010,Maali2012,Lei2016}.
From the theoretical side, the pioneering work by Bocquet and Barrat \cite{Bocquet1993,Bocquet1994} showed that $\lambda$ 
can be related to the equilibrium fluctuations of the friction force through a Green-Kubo (GK) formula: 
\begin{equation}\label{eq:BB}
	\lambda = %
	\lim_{t \to \infty}
	\frac{1}{S \kB T}
	\int_0^t 
	\bracket{%
	\delta F(t')\,\delta F(0)
	} \mathrm{d}t',
\end{equation} 
where $\delta F$ is the random friction force on the wall at equilibrium, $S$ the wall surface area and $T$ the temperature, with $k_\text{B}$ the Boltzmann constant. Later, the system size dependence of the formula, called the \textit{plateau problem}, was pointed out \cite{Petravic2007} and alternative ways to estimate the FC have been proposed \cite{Sokhan2008,Hansen2011,Huang2014,Oga2019}. 

Recently, two theoretical approaches have been proposed to challenge the plateau problem fundamentally. Espa\~nol and coworkers \cite{Espanol2019,DeLaTorre2019} developed a new theory of non-equilibrium statistical mechanics, 
which led to a corrected form of the GK formula under the assumption that the system is Markovian. 
In another recent work, \citet{Nakano2019a,Nakano2019c} introduced explicit assumptions on the scale separation between the microscopic motion of molecules and the macroscopic motion of fluid and proposed a new way to estimate $\lambda$ based on linearized fluctuating hydrodynamics.
Both works from the two groups involved elaborate mathematical manipulations, and only reported the pure viscous (Markovian) behavior of the Navier FC, for a Lennard-Jones (LJ) liquid on a simple model wall. 
However, non-Markovian behavior of the FC was recently reported for a LJ liquid on a fcc lattice \cite{Omori2019a} and it is plausible that more complex liquids such as water also show such behavior, in analogy with their bulk transport properties \cite{slie1966,masciovecchio2004,Omelyan2005,osullivan2019,Straube2020,Schulz2020}. 

In this Letter, we develop a theory to relate the Navier FC and the random force autocorrelation on the wall, by employing rather classical tools such as Stokes and Langevin equations. 
This theoretical framework offers some new perspectives on long-standing debates related to the GK modeling of liquid-solid friction, together with a simple and fast method to fully characterize the frequency-dependent Navier FC. We then apply this method to explore the frictional behavior of a simple LJ liquid and of water at various temperatures -- including the supercooled regime.

\begin{figure}
\centering
\includegraphics[width=\linewidth]{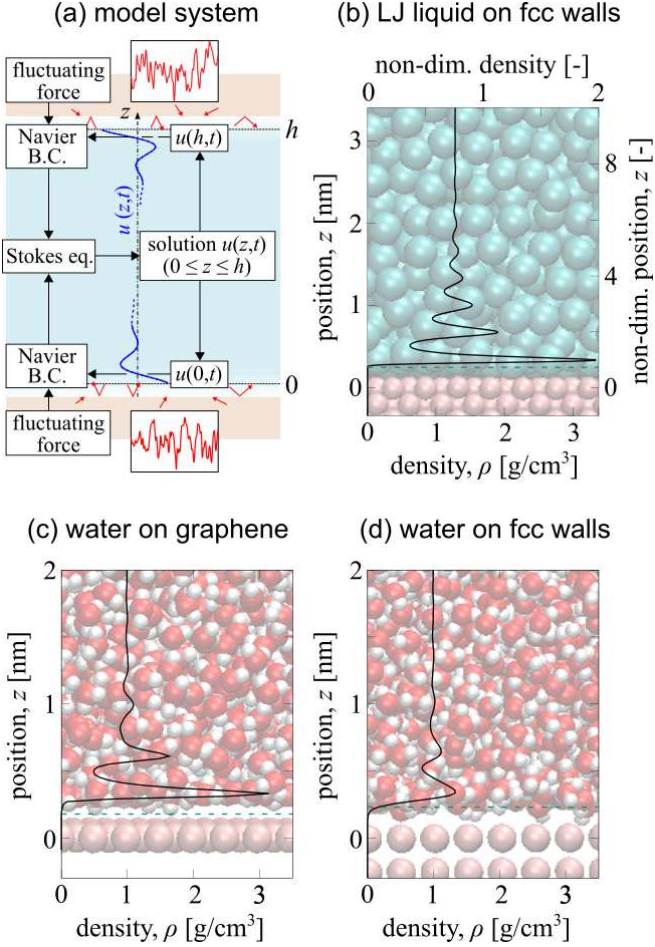}
\caption{(a) Model system for the derivation of the Navier friction coefficient, and (b-d) snapshots of the bottom half of the systems tested in this study overlaid with density distributions. The dotted green lines denote the position of the hydrodynamic boundaries. For (b), the system with a contact angle of 79 degrees is shown. We give the details on the system dimensions in the supplemental material.} \label{fig:schematics}
\end{figure}

\textit{Theory}\textemdash
Let us consider the system shown in Fig.~\ref{fig:schematics}(a), where a liquid is confined between two solid walls under no external field. 
When the bottom wall is let to move freely in a wall-tangential direction $x$, its motion can be described by a Langevin equation \cite{Bocquet2013}:
\begin{equation}\label{eq:langevin}
	M\dd{U}{t}=-S\int_0^t\xi(t-t')U(t') \rd t'+\delta F^{\scrm{bot}},
\end{equation}
where $M$, $S$ and $U$ are the mass, the surface area and the $x$-direction velocity of the bottom wall, respectively; $\xi$ is the friction kernel and $\delta F^{\scrm{bot}}$ is the random force that originates from the direct interaction between the solid and liquid particles. Assuming the equipartition of energy, Eq.~\eqref{eq:langevin} leads to the fluctuation-dissipation theorem: 
\begin{equation}\label{eq:rand}
	C_{\delta F^{\scrm{bot}}}(t) \coloneqq
	\bracket{%
	\delta F^{\scrm{bot}}(t)\,
	\delta F^{\scrm{bot}}(0)
	} = S \kB T \xi(t).
\end{equation} 
The motion of the liquid in response to the bottom wall motion (the top wall is fixed) can be described by Stokes equation for a wide frequency range \cite{Omori2019a} \footnote{As we show in the supplemental material, the frequency response characteristics of the FC and of the bulk viscosity are similar. This means that there is no bulk flow in the time scales where the high-frequency mode of the FC such as the elasticity dominates \cite{Omori2019a}, and therefore one can neglect the frequency-dependent component of the bulk viscosity in the present analysis.}:
\begin{equation}\label{eq:stokes}
	\pd{u(z,t)}{t}=\frac{\eta}{\rho}\pdd{u(z,t)}{z},
\end{equation}
with the Navier boundary condition defined on the bottom and top hydrodynamic boundaries
\begin{equation}\label{eq:bc}
	\begin{cases}
		\eta\left.\pd{u(z,t)}{z}\right|_{z=0}=%
		\int_0^t \lambda(t-t')\left[
		u(0,t')-U(t')
		\right]\rd t' , \\[2mm]
		\eta\left.\pd{u(z,t)}{z}\right|_{z=h}=%
		\int_0^t \lambda(t-t')\left[
		-u(h,t')
		\right]\rd t',
	\end{cases}
\end{equation}
where $u$, $t$, $\rho$, $\eta$ and $\lambda$  denote the liquid velocity in the $x$ direction, the time, the bulk liquid density, the bulk liquid viscosity, and the Navier FC, respectively. Note that $\lambda$ is frequency dependent and of non-Markovian nature: $\lambda$ has the dimension [Pa/m] instead of $[\mathrm{Pa}\cdot \mathrm{s/m}]$ adequate for the frequency-independent FC. 
Because the first term on the RHS of Eq.~\eqref{eq:langevin} would also be written as $-S\int_0^t \lambda(t-t')[U(t')-u(0,t')]\rd t'$ in terms of the slip velocity on the wall, the friction kernel $\xi$ is given 
from the solution of Eqs.~\eqref{eq:stokes} and \eqref{eq:bc} (for the complete derivation, see the supplemental material, SM). Combined with Eq.~\eqref{eq:rand}, the expression for the force autocorrelation function is obtained: 
\begin{equation}\label{eq:bot}
	\frac{\fl{C}_{\delta F^{\scrm{bot}}}}{S \kB T}
	=%
	\frac{%
	\fl{\lambda} \eta\zeta
	\left[
	\eta\zeta \sinh(\zeta h) + 
	\fl{\lambda} \cosh(\zeta h)
	\right]
	}
	{%
	(\fl{\lambda}^2 + \eta^2 \zeta^2)\sinh(\zeta h) + 
	2 \fl{\lambda}\eta\zeta \cosh(\zeta h)
	},
\end{equation}
where the tilde indicates that the variables are Fourier-Laplace transformed and $\zeta$ denotes $\sqrt{i\rho\omega/\eta}$, with $\omega$ the angular frequency. 
Considering that the nature of the random force is independent of the wall velocity by construction, Eq.~\eqref{eq:bot} holds even when the bottom wall is fixed.
Unlike the GK formula by \citet{Bocquet2013}, this equation explicitly includes the system size dependence in the relation between the random force autocorrelation and the Navier FC. 
It is also more general as it provides the viscoelastic behavior of the friction coefficient. 
From Eq.~\eqref{eq:bot}, the asymptotic behaviors of the random force autocorrelation are
\begin{eqnarray}
	\lim_{\omega \to 0,\;h\scrm{:finite}}%
	\frac{\fl{C}_{\delta F^{\scrm{bot}}}}{S \kB T}
	&=&
	\frac{\lambda_0}{h/b+2}\label{eq:asymp_omega}\quad \mathrm{and}\\
	\lim_{h \to \infty,\;\omega\scrm{:finite}}%
	\frac{\fl{C}_{\delta F^{\scrm{bot}}}}{S \kB T}
	&=&
	\frac{\fl{\lambda}}{\fl{\lambda}/(\eta\zeta)+1},\label{eq:asymp_h}
\end{eqnarray}
where $\lambda_0$ is the zero-frequency component of the Navier FC 
and $b$ is the slip length defined as $\eta/\lambda_0$. 

From Eqns.~\eqref{eq:asymp_omega} and \eqref{eq:asymp_h}, one can derive several important properties of the random force, whose evidence will be shown later in the \textit{Results} section. First, the GK integral of the random force (Eq.~\ref{eq:asymp_omega}) is not zero but has a finite value that depends on the system height $h$
, which tells that the integral \textit{has} a plateau. 
Note that the Bocquet-Barrat formula may be recovered by taking the plug-flow limit $b \gg h$. In this limit only, the GK integral no longer depends on the system height $h$. 
Second, in the thermodynamic limit where the system height is infinite, the GK integral of the random force goes to zero: this is true regardless of the order in which the limits are taken, %
$\lim_{h \to \infty}\lim_{\omega \to 0}\fl{C}_{\delta F^{\scrm{bot}}}=%
\lim_{\omega \to 0}\lim_{h \to \infty}\fl{C}_{\delta F^{\scrm{bot}}}=0$. Finally, when the frequency $\omega$ is high enough so that the penetration length is much smaller than the magnitude of the complex slip length ($\sqrt{\eta/\rho\omega} \ll |\eta/\fl{\lambda}|$) as well as than the system height ($\sqrt{\eta/\rho\omega} \ll h$), the random force autocorrelation coincides with the Navier FC:
\begin{equation}\label{eq:small_t}
	\frac{%
	C_{\delta F^{\scrm{bot}}}(t)
	}
	{S \kB T}
	\approx \lambda(t)
\end{equation}
for small $t$ satisfying $\sqrt{\eta t/\rho} \ll \min\{|\eta/\fl{\lambda}|,h\}$. All these results, which are in contrast to the common view that $\lambda_0$ might be obtained as $\lim_{\omega \to 0}\lim_{h \to \infty}\fl{C}_{\delta F^{\scrm{bot}}}$ \cite{Mazur1970,Bocquet2013}, reflect our explicit consideration of the system height and its hydrodynamic influence on the fluctuations of the friction force.

Interestingly, one can find the Navier FC for the whole frequency range from the measured random force autocorrelation $C_{\delta F^{\scrm{bot}}}$ by solving Eq.~\eqref{eq:bot} for $\fl{\lambda}$, which is a quadratic equation of it. The physically correct solution out of the two can be chosen so that it satisfies the following relationship:
\begin{equation}\label{eq:total}
	\frac{\fl{C}_{\delta F^{\scrm{total}}}}{2S \kB T}
	=%
	\frac{%
	\fl{\lambda} \eta\zeta 
	(\eta\zeta \sinh(\zeta h) +
	\fl{\lambda}(\cosh(\zeta h)-1))
	}
	{%
	(\fl{\lambda}^2 + \eta^2 \zeta^2)\sinh(\zeta h) + 
	2 \fl{\lambda}\eta\zeta \cosh(\zeta h)
	},
\end{equation}
where $C_{\delta F^{\scrm{total}}}$ is the autocorrelation function of the random force summed over both the top and bottom walls (for the complete procedure, see the SM). 

\textit{Simulation}\textemdash
To validate the ideas in the \textit{Theory} section, we performed equilibrium molecular dynamics (MD) simulations for two kinds of liquid on different walls.  

The first system was a LJ liquid confined between two fcc crystal walls (Fig.~\ref{fig:schematics}b). The liquid consisted of 6400 molecules unless otherwise mentioned. The quantities for this system are shown in LJ reduced units based on the liquid molecular mass $m_\mathrm{f}$ ($=6.634 \times 10^{-26}\ \mathrm{kg}$) and two parameters $\ff{\sigma}$ ($=3.40$ \AA) and $\ff{\varepsilon}$ ($=1.67 \times 10^{-21}$ J) for the LJ potential 
$\Phi_{\scrm{LJ}}(r_{ij}) 
=4\varepsilon \left[ \left( \sigma / r_{ij} \right)^{12} -\left( \sigma / r_{ij} \right)^6 \right]$
between the liquid molecules, with $r_{ij}$ the distance between particle $i$ and $j$ with a cut-off distance of $3.50\sigma$ \cite{Nishida2014}. 
Each solid wall consisted of 8 layers of atoms in the (001) plane of a fcc crystal with a lattice constant of $1.15\ff{\sigma}$. 
%
The system temperature was controlled at $0.827\ff{\varepsilon}/\kB$ by a Langevin thermostat set on the second outermost layer of the walls, and the pressure was set to $0.094\ff{\varepsilon}/\ff{\sigma}^3$ by a preliminary piston equilibration (see Ref.~\citenum{Omori2019a} for technical details). 
For the liquid-solid interaction, the LJ potential was adopted as well.
To see the effect of the wettability, which is known to have impact on the friction \cite{Bocquet2010}, three different
$\fw{\varepsilon}$ were used, $\fw{\varepsilon}=\fw{\varepsilon}^{0}=0.155\ff{\varepsilon}$,  $\fw{\varepsilon}=2\fw{\varepsilon}^{0}$, and $\fw{\varepsilon}=3\fw{\varepsilon}^{0}$, while $\fw{\sigma}=1.01\ff{\sigma}$ was kept constant. 
The corresponding contact angles of a sessile LJ droplet on the three
walls 
were 136, 79 and 0 degrees, respectively \cite{Ogawa2019}.

The other two systems were water confined between either graphene walls or fcc crystal walls (Figs.~\ref{fig:schematics}c and \ref{fig:schematics}d, respectively).
In this case the fluid was constituted by 4096 TIP4P/2005 water molecules \cite{abascal2005}. For water enclosed between generic fcc walls, such walls were constituted by three atomic layers of a fcc crystal exposing the (001) face
with a lattice constant of $5.356\,$\AA{}, and with 
liquid-solid interaction parameters taken for hydrophobic walls from Ref.~\citenum{Huang2008}. For water enclosed between graphene walls, the liquid-solid interaction parameters were taken from Ref.~\citenum{falk2010}. For both systems, the temperature was controlled by applying a Nos\'e{}-Hoover thermostat to liquid atoms, and the pressure was set to $1\,$atm through a preliminary piston equilibration (see Ref.~\citenum{Herrero2020} for technical details).

\begin{figure}
\centering
\includegraphics[width=1.0\linewidth]{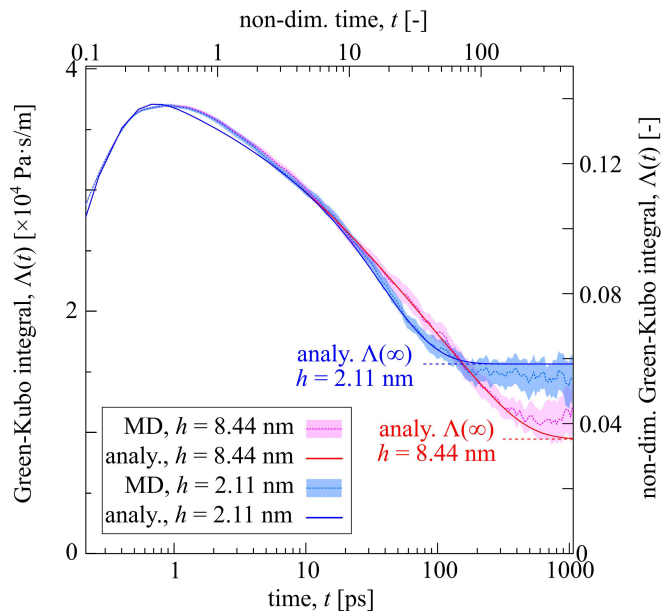}
\caption{Size dependence of the finite-time GK integral for the system 
in Fig.~1(b):
comparison between analytical predictions and MD results for two system heights $h$. The liquid consisted of 6400 and 1600 molecules for $h=8.44\ \mathrm{nm}$ and $2.11\ \mathrm{nm}$, respectively.}
\label{fig:GK}
\end{figure}

\textit{Results}\textemdash
First, we discuss the convergence of the GK integral of the random force autocorrelation. 
Figure~\ref{fig:GK} shows the normalized finite-time GK integral, $\Lambda(t) \coloneqq \int_0^t C_{\delta F^{\scrm{bot}}}(t') \rd t'/ S \kB T$,
as a function of the upper time limit of the integration, for the LJ liquid confined by walls with 
$\fw{\varepsilon}= 2 \varepsilon^{0}_\mathrm{fw}$.
To obtain $\Lambda(t)$ up to $t=1000$ ps, we produced the simulation data typically for 300\,ns.
One can see that $\Lambda(t)$ has a system-height-dependent plateau for $t\to\infty$, whose value, the GK integral, decreases by increasing the system height. The figure also illustrates that the whole $\Lambda(t)$ profile is well reproduced with $C_{\delta F^{\scrm{bot}}}$ calculated from the RHS of Eq.~\eqref{eq:bot},
and the plateau values coincide with the RHS of Eq.~\eqref{eq:asymp_omega}. In this evaluation of Eqs.~\eqref{eq:bot} and \eqref{eq:asymp_omega}, we substituted $\lambda(t)$ by the Maxwell-type model $\lambda_0 \exp (-t/t_{\lambda})/t_{\lambda}$ with the parameters $\lambda_0=0.1492\sqrt{m_\mathrm{f}\ff{\varepsilon}}/\ff{\sigma}^3$ and $t_{\lambda}=0.077\ff{\sigma}\sqrt{m_\mathrm{f}/\ff{\varepsilon}}$ taken from the results of non-equilibrium simulations \cite{Omori2019a}, whose simulation system and conditions were identical to the present study \footnote{The dimensions of the larger system was identical to Ref.~\citenum{Omori2019a}.  Therefore, the hydrodynamics height of the system $h$ of the larger system was taken from Ref.~\citenum{Omori2019a}. For the smaller system, $h$ was calculated considering that the hydrodynamic wall position should be identical to that of the larger system.}. Here, the same $\lambda$ was used regardless of the system height: this shows that there is no system size dependence in the estimation of $\lambda$ from the MD simulation data by the present theory. 
Sometimes in the literature \cite{Bocquet1994,Joly2016}, 
$\lambda_0$ is estimated as $\max\Lambda(t)$, which gives about $0.14\sqrt{m_\mathrm{f}\ff{\varepsilon}}/\ff{\sigma}^3$ and slightly underestimates $\lambda_0$. This under-estimation was also shown in Ref.~\citenum{Oga2019}. 
In summary, the discussion here provides a new perspective to the long-standing plateau issue for the evaluation of the GK integral: there is a plateau in the limit of the integral for finite-sized systems and this plateau value has a hydrodynamic meaning.

Now we determine the complete nature of the Navier FC for the three liquid-solid interfaces shown in Fig.~\ref{fig:schematics}. 
As described in the \textit{Theory} section, the Navier FC $\lambda$ can be obtained by solving Eq.~\eqref{eq:bot} for $\lambda$ once the bulk liquid properties and the random force autocorrelation are measured. To estimate the hydrodynamic system height $h$, for the LJ liquid system we employed the values from Ref.~\citenum{Omori2019a} and for the water systems we adopted the separation between the Gibbs dividing surfaces \cite{Herrero2019} on the two confining walls. 
\begin{figure*}
\centering
\includegraphics[width=\linewidth]{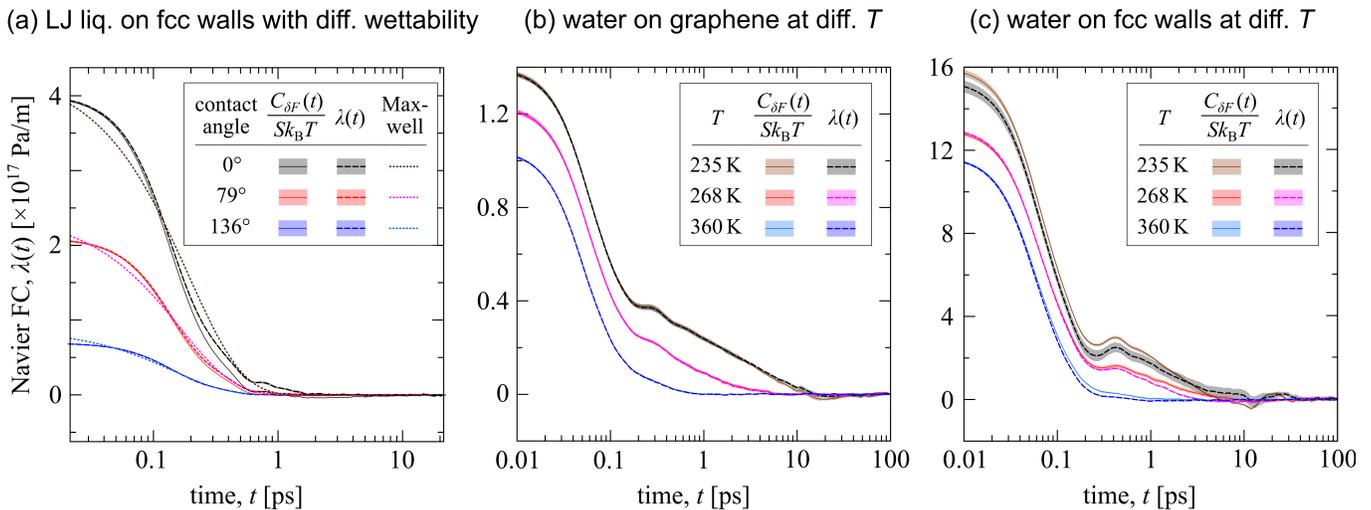}
\caption{Random force autocorrelation function $C_{\delta F^{\scrm{bot}}}(t) / (S \kB T)$ and friction coefficient $\lambda(t)$ obtained from Eq.~(\ref{eq:bot}) for three different systems. The Maxwell viscoelastic model is also shown for (a).} 
\label{fig:lambda}
\end{figure*}

Figure~\ref{fig:lambda} shows the Navier FC $\lambda(t)$ and the normalized random force autocorrelation $C_{\delta F^{\scrm{bot}}}(t) / S \kB T$ for the LJ liquid and for water. 
The equivalence between $\lambda$ and $C_{\delta F^{\scrm{bot}}} / S \kB T$, Eq.~\eqref{eq:small_t}, holds almost everywhere, although the condition $\sqrt{\eta t/\rho} \ll \min\{|\eta/\fl{\lambda}|,h\}$ is not strictly satisfied in the tail region for supercooled water.
Figure~\ref{fig:lambda}(a) shows $\lambda(t)$ for the LJ liquid system, together with a Maxwell-type viscoelastic model $\lambda_M(t)=\lambda_0 \exp (-t/t_{\lambda})/t_{\lambda}$ \cite{Omori2019a} for comparison. 
The Maxwell-type model describes well the time-dependent behavior of the FC for this simple liquid, which was expected from Fig.~\ref{fig:GK} showing good reproduction of the GK integral applying $\lambda(t)=\lambda_M(t)$ in  Eq.~\eqref{eq:bot}. A slight distortion in the $\lambda_M$ profile comes from the non-vanishing time derivative at $t=0$. Note that $\rd \lambda / \rd t|_{t=0}=0$ follows from Eq.~\eqref{eq:small_t} because the autocorrelation function is an even function in a stationary system.
Figures~\ref{fig:lambda}(b) and (c) show $\lambda(t)$ for water confined either by graphene walls or by \green{fcc} walls at three different temperatures: 235\,K, 268\,K and 360\,K. The profiles of the FC on both walls at 360\,K (i.e. for liquid water above its melting point) look similar to those of the LJ liquid. 
However, for supercooled water, i.e. metastable liquid water below its melting point, $\lambda(t)$ cannot be described by the Maxwell-type model. It is known for supercooled liquids that the density relaxation is a two-step process with two characteristic decay times \cite{kob1995,Gallo1996}: here one can see that the FC also decays with two characteristic times.


\textit{Conclusions}\textemdash
We have derived analytical expressions to connect the equilibrium fluctuations of the random force on the wall and the Navier friction coefficient (FC). The expressions are generic in the sense that they explicitly include the system size dependence and also the FC can be frequency dependent, which enabled us to address (i) the plateau issue on the evaluation of the FC and (ii) the non-Markovian behavior of the liquid-solid friction.  For (i) we found that the Green-Kubo integral of the random force autocorrelation has actually a plateau for the finite-sized systems and the plateau value has a clear hydrodynamic meaning. For (ii) we evaluated the frequency-/time-dependent FC from equilibrium molecular dynamics simulation data for a Lennard-Jones (LJ) liquid and for water under different wall confinements and temperatures, without ambiguity due to the simulation system size. We showed that the Maxwell viscoelastic model is a fair approximation for the FC of LJ liquid on a fcc wall, and similarly a model with a single relaxation time can be applied for water on fcc and graphene walls at a high temperature, but a model with more than one time scale is required to describe the FC of supercooled water on both walls.
Our theoretical  framework opens the way to explore the frequency dependent FC for a wide range of complex liquids by non-demanding atomistic simulations, whose system size may be small.

\begin{acknowledgments}
This work was financially supported by JSPS KAKENHI Grant Nos.\ 18K03929 and 18K03978, and by the ANR, Project ANR-16-CE06-0004-01 NECtAR. YY was also supported by JST CREST Grant No.\ JPMJCR18I1, Japan. LJ was supported by the Institut Universitaire de France.
\end{acknowledgments}


\begin{thebibliography}{46}%
\makeatletter
\providecommand \@ifxundefined [1]{%
 \@ifx{#1\undefined}
}%
\providecommand \@ifnum [1]{%
 \ifnum #1\expandafter \@firstoftwo
 \else \expandafter \@secondoftwo
 \fi
}%
\providecommand \@ifx [1]{%
 \ifx #1\expandafter \@firstoftwo
 \else \expandafter \@secondoftwo
 \fi
}%
\providecommand \natexlab [1]{#1}%
\providecommand \enquote  [1]{``#1''}%
\providecommand \bibnamefont  [1]{#1}%
\providecommand \bibfnamefont [1]{#1}%
\providecommand \citenamefont [1]{#1}%
\providecommand \href@noop [0]{\@secondoftwo}%
\providecommand \href [0]{\begingroup \@sanitize@url \@href}%
\providecommand \@href[1]{\@@startlink{#1}\@@href}%
\providecommand \@@href[1]{\endgroup#1\@@endlink}%
\providecommand \@sanitize@url [0]{\catcode `\\12\catcode `\$12\catcode
  `\&12\catcode `\#12\catcode `\^12\catcode `\_12\catcode `\%12\relax}%
\providecommand \@@startlink[1]{}%
\providecommand \@@endlink[0]{}%
\providecommand \url  [0]{\begingroup\@sanitize@url \@url }%
\providecommand \@url [1]{\endgroup\@href {#1}{\urlprefix }}%
\providecommand \urlprefix  [0]{URL }%
\providecommand \Eprint [0]{\href }%
\providecommand \doibase [0]{https://doi.org/}%
\providecommand \selectlanguage [0]{\@gobble}%
\providecommand \bibinfo  [0]{\@secondoftwo}%
\providecommand \bibfield  [0]{\@secondoftwo}%
\providecommand \translation [1]{[#1]}%
\providecommand \BibitemOpen [0]{}%
\providecommand \bibitemStop [0]{}%
\providecommand \bibitemNoStop [0]{.\EOS\space}%
\providecommand \EOS [0]{\spacefactor3000\relax}%
\providecommand \BibitemShut  [1]{\csname bibitem#1\endcsname}%
\let\auto@bib@innerbib\@empty
\bibitem [{\citenamefont {Sparreboom}\ \emph {et~al.}(2009)\citenamefont
  {Sparreboom}, \citenamefont {van~den Berg},\ and\ \citenamefont
  {Eijkel}}]{Sparreboom2009}%
  \BibitemOpen
  \bibfield  {author} {\bibinfo {author} {\bibfnamefont {W.}~\bibnamefont
  {Sparreboom}}, \bibinfo {author} {\bibfnamefont {A.}~\bibnamefont {van~den
  Berg}},\ and\ \bibinfo {author} {\bibfnamefont {J.~C.~T.}\ \bibnamefont
  {Eijkel}},\ }\bibfield  {title} {\bibinfo {title} {{Principles and
  applications of nanofluidic transport}},\ }\href
  {https://doi.org/10.1038/nnano.2009.332} {\bibfield  {journal} {\bibinfo
  {journal} {Nat. Nanotechnol.}\ }\textbf {\bibinfo {volume} {4}},\ \bibinfo
  {pages} {713} (\bibinfo {year} {2009})}\BibitemShut {NoStop}%
\bibitem [{\citenamefont {Bocquet}\ and\ \citenamefont
  {Charlaix}(2010)}]{Bocquet2010}%
  \BibitemOpen
  \bibfield  {author} {\bibinfo {author} {\bibfnamefont {L.}~\bibnamefont
  {Bocquet}}\ and\ \bibinfo {author} {\bibfnamefont {E.}~\bibnamefont
  {Charlaix}},\ }\bibfield  {title} {\bibinfo {title} {{Nanofluidics, from bulk
  to interfaces}},\ }\href {https://doi.org/10.1039/B909366B} {\bibfield
  {journal} {\bibinfo  {journal} {Chem. Soc. Rev.}\ }\textbf {\bibinfo {volume}
  {39}},\ \bibinfo {pages} {1073} (\bibinfo {year} {2010})}\BibitemShut
  {NoStop}%
\bibitem [{\citenamefont {Daiguji}(2010)}]{Daiguji2010}%
  \BibitemOpen
  \bibfield  {author} {\bibinfo {author} {\bibfnamefont {H.}~\bibnamefont
  {Daiguji}},\ }\bibfield  {title} {\bibinfo {title} {{Ion transport in
  nanofluidic channels}},\ }\href@noop {} {\bibfield  {journal} {\bibinfo
  {journal} {Chem. Soc. Rev.}\ }\textbf {\bibinfo {volume} {39}},\ \bibinfo
  {pages} {901} (\bibinfo {year} {2010})}\BibitemShut {NoStop}%
\bibitem [{\citenamefont {Faucher}\ \emph {et~al.}(2019)\citenamefont
  {Faucher}, \citenamefont {Aluru}, \citenamefont {Bazant}, \citenamefont
  {Blankschtein}, \citenamefont {Brozena}, \citenamefont {Cumings},
  \citenamefont {Pedro~de Souza}, \citenamefont {Elimelech}, \citenamefont
  {Epsztein}, \citenamefont {Fourkas}, \citenamefont {Rajan}, \citenamefont
  {Kulik}, \citenamefont {Levy}, \citenamefont {Majumdar}, \citenamefont
  {Martin}, \citenamefont {McEldrew}, \citenamefont {Misra}, \citenamefont
  {Noy}, \citenamefont {Pham}, \citenamefont {Reed}, \citenamefont {Schwegler},
  \citenamefont {Siwy}, \citenamefont {Wang},\ and\ \citenamefont
  {Strano}}]{Faucher2019}%
  \BibitemOpen
  \bibfield  {author} {\bibinfo {author} {\bibfnamefont {S.}~\bibnamefont
  {Faucher}}, \bibinfo {author} {\bibfnamefont {N.}~\bibnamefont {Aluru}},
  \bibinfo {author} {\bibfnamefont {M.~Z.}\ \bibnamefont {Bazant}}, \bibinfo
  {author} {\bibfnamefont {D.}~\bibnamefont {Blankschtein}}, \bibinfo {author}
  {\bibfnamefont {A.~H.}\ \bibnamefont {Brozena}}, \bibinfo {author}
  {\bibfnamefont {J.}~\bibnamefont {Cumings}}, \bibinfo {author} {\bibfnamefont
  {J.}~\bibnamefont {Pedro~de Souza}}, \bibinfo {author} {\bibfnamefont
  {M.}~\bibnamefont {Elimelech}}, \bibinfo {author} {\bibfnamefont
  {R.}~\bibnamefont {Epsztein}}, \bibinfo {author} {\bibfnamefont {J.~T.}\
  \bibnamefont {Fourkas}}, \bibinfo {author} {\bibfnamefont {A.~G.}\
  \bibnamefont {Rajan}}, \bibinfo {author} {\bibfnamefont {H.~J.}\ \bibnamefont
  {Kulik}}, \bibinfo {author} {\bibfnamefont {A.}~\bibnamefont {Levy}},
  \bibinfo {author} {\bibfnamefont {A.}~\bibnamefont {Majumdar}}, \bibinfo
  {author} {\bibfnamefont {C.}~\bibnamefont {Martin}}, \bibinfo {author}
  {\bibfnamefont {M.}~\bibnamefont {McEldrew}}, \bibinfo {author}
  {\bibfnamefont {R.~P.}\ \bibnamefont {Misra}}, \bibinfo {author}
  {\bibfnamefont {A.}~\bibnamefont {Noy}}, \bibinfo {author} {\bibfnamefont
  {T.~A.}\ \bibnamefont {Pham}}, \bibinfo {author} {\bibfnamefont
  {M.}~\bibnamefont {Reed}}, \bibinfo {author} {\bibfnamefont {E.}~\bibnamefont
  {Schwegler}}, \bibinfo {author} {\bibfnamefont {Z.}~\bibnamefont {Siwy}},
  \bibinfo {author} {\bibfnamefont {Y.}~\bibnamefont {Wang}},\ and\ \bibinfo
  {author} {\bibfnamefont {M.}~\bibnamefont {Strano}},\ }\bibfield  {title}
  {\bibinfo {title} {{Critical Knowledge Gaps in Mass Transport through
  Single-Digit Nanopores: A Review and Perspective}},\ }\href
  {https://doi.org/10.1021/acs.jpcc.9b02178} {\bibfield  {journal} {\bibinfo
  {journal} {J. Phys. Chem. C}\ }\textbf {\bibinfo {volume} {123}},\ \bibinfo
  {pages} {21309} (\bibinfo {year} {2019})}\BibitemShut {NoStop}%
\bibitem [{\citenamefont {Kavokine}\ \emph {et~al.}(2021)\citenamefont
  {Kavokine}, \citenamefont {Netz},\ and\ \citenamefont
  {Bocquet}}]{Kavokine2021}%
  \BibitemOpen
  \bibfield  {author} {\bibinfo {author} {\bibfnamefont {N.}~\bibnamefont
  {Kavokine}}, \bibinfo {author} {\bibfnamefont {R.~R.}\ \bibnamefont {Netz}},\
  and\ \bibinfo {author} {\bibfnamefont {L.}~\bibnamefont {Bocquet}},\
  }\bibfield  {title} {\bibinfo {title} {{Fluids at the Nanoscale: From
  Continuum to Subcontinuum Transport}},\ }\href
  {https://doi.org/10.1146/annurev-fluid-071320-095958} {\bibfield  {journal}
  {\bibinfo  {journal} {Ann. Rev. Fluid Mech.}\ }\textbf {\bibinfo {volume}
  {53}},\ \bibinfo {pages} {377} (\bibinfo {year} {2021})}\BibitemShut
  {NoStop}%
\bibitem [{\citenamefont {Bocquet}(2020)}]{Bocquet2020}%
  \BibitemOpen
  \bibfield  {author} {\bibinfo {author} {\bibfnamefont {L.}~\bibnamefont
  {Bocquet}},\ }\bibfield  {title} {\bibinfo {title} {{Nanofluidics coming of
  age}},\ }\href {https://doi.org/10.1038/s41563-020-0625-8} {\bibfield
  {journal} {\bibinfo  {journal} {Nat. Mater.}\ }\textbf {\bibinfo {volume}
  {19}},\ \bibinfo {pages} {254} (\bibinfo {year} {2020})}\BibitemShut
  {NoStop}%
\bibitem [{\citenamefont {Joly}\ \emph {et~al.}(2004)\citenamefont {Joly},
  \citenamefont {Ybert}, \citenamefont {Trizac},\ and\ \citenamefont
  {Bocquet}}]{Joly2004}%
  \BibitemOpen
  \bibfield  {author} {\bibinfo {author} {\bibfnamefont {L.}~\bibnamefont
  {Joly}}, \bibinfo {author} {\bibfnamefont {C.}~\bibnamefont {Ybert}},
  \bibinfo {author} {\bibfnamefont {E.}~\bibnamefont {Trizac}},\ and\ \bibinfo
  {author} {\bibfnamefont {L.}~\bibnamefont {Bocquet}},\ }\bibfield  {title}
  {\bibinfo {title} {{Hydrodynamics within the electric double layer on
  slipping surfaces}},\ }\href {https://doi.org/10.1103/PhysRevLett.93.257805}
  {\bibfield  {journal} {\bibinfo  {journal} {Phys. Rev. Lett.}\ }\textbf
  {\bibinfo {volume} {93}},\ \bibinfo {pages} {257805} (\bibinfo {year}
  {2004})}\BibitemShut {NoStop}%
\bibitem [{\citenamefont {Ajdari}\ and\ \citenamefont
  {Bocquet}(2006)}]{Ajdari2006}%
  \BibitemOpen
  \bibfield  {author} {\bibinfo {author} {\bibfnamefont {A.}~\bibnamefont
  {Ajdari}}\ and\ \bibinfo {author} {\bibfnamefont {L.}~\bibnamefont
  {Bocquet}},\ }\bibfield  {title} {\bibinfo {title} {{Giant Amplification of
  Interfacially Driven Transport by Hydrodynamic Slip: Diffusio-Osmosis and
  Beyond}},\ }\href {https://doi.org/10.1103/PhysRevLett.96.186102} {\bibfield
  {journal} {\bibinfo  {journal} {Phys. Rev. Lett.}\ }\textbf {\bibinfo
  {volume} {96}},\ \bibinfo {pages} {186102} (\bibinfo {year}
  {2006})}\BibitemShut {NoStop}%
\bibitem [{\citenamefont {Ren}\ and\ \citenamefont {Stein}(2008)}]{Ren2008}%
  \BibitemOpen
  \bibfield  {author} {\bibinfo {author} {\bibfnamefont {Y.}~\bibnamefont
  {Ren}}\ and\ \bibinfo {author} {\bibfnamefont {D.}~\bibnamefont {Stein}},\
  }\bibfield  {title} {\bibinfo {title} {{Slip-enhanced electrokinetic energy
  conversion in nanofluidic channels}},\ }\href
  {https://doi.org/10.1088/0957-4484/19/19/195707} {\bibfield  {journal}
  {\bibinfo  {journal} {Nanotechnology}\ }\textbf {\bibinfo {volume} {19}},\
  \bibinfo {pages} {195707} (\bibinfo {year} {2008})}\BibitemShut {NoStop}%
\bibitem [{\citenamefont {Radha}\ \emph {et~al.}(2016)\citenamefont {Radha},
  \citenamefont {Esfandiar}, \citenamefont {Wang}, \citenamefont {Rooney},
  \citenamefont {Gopinadhan}, \citenamefont {Keerthi}, \citenamefont
  {Mishchenko}, \citenamefont {Janardanan}, \citenamefont {Blake},
  \citenamefont {Fumagalli}, \citenamefont {Lozada-Hidalgo}, \citenamefont
  {Garaj}, \citenamefont {Haigh}, \citenamefont {Grigorieva}, \citenamefont
  {Wu},\ and\ \citenamefont {Geim}}]{Radha2016}%
  \BibitemOpen
  \bibfield  {author} {\bibinfo {author} {\bibfnamefont {B.}~\bibnamefont
  {Radha}}, \bibinfo {author} {\bibfnamefont {A.}~\bibnamefont {Esfandiar}},
  \bibinfo {author} {\bibfnamefont {F.~C.}\ \bibnamefont {Wang}}, \bibinfo
  {author} {\bibfnamefont {A.~P.}\ \bibnamefont {Rooney}}, \bibinfo {author}
  {\bibfnamefont {K.}~\bibnamefont {Gopinadhan}}, \bibinfo {author}
  {\bibfnamefont {A.}~\bibnamefont {Keerthi}}, \bibinfo {author} {\bibfnamefont
  {A.}~\bibnamefont {Mishchenko}}, \bibinfo {author} {\bibfnamefont
  {A.}~\bibnamefont {Janardanan}}, \bibinfo {author} {\bibfnamefont
  {P.}~\bibnamefont {Blake}}, \bibinfo {author} {\bibfnamefont
  {L.}~\bibnamefont {Fumagalli}}, \bibinfo {author} {\bibfnamefont
  {M.}~\bibnamefont {Lozada-Hidalgo}}, \bibinfo {author} {\bibfnamefont
  {S.}~\bibnamefont {Garaj}}, \bibinfo {author} {\bibfnamefont {S.~J.}\
  \bibnamefont {Haigh}}, \bibinfo {author} {\bibfnamefont {I.~V.}\ \bibnamefont
  {Grigorieva}}, \bibinfo {author} {\bibfnamefont {H.~A.}\ \bibnamefont {Wu}},\
  and\ \bibinfo {author} {\bibfnamefont {A.~K.}\ \bibnamefont {Geim}},\
  }\bibfield  {title} {\bibinfo {title} {{Molecular transport through
  capillaries made with atomic-scale precision}},\ }\href
  {https://doi.org/10.1038/nature19363} {\bibfield  {journal} {\bibinfo
  {journal} {Nature}\ }\textbf {\bibinfo {volume} {538}},\ \bibinfo {pages}
  {222} (\bibinfo {year} {2016})}\BibitemShut {NoStop}%
\bibitem [{\citenamefont {Navier}(1823)}]{Navier1823}%
  \BibitemOpen
  \bibfield  {author} {\bibinfo {author} {\bibfnamefont {C.}~\bibnamefont
  {Navier}},\ }\bibfield  {title} {\bibinfo {title} {{M{\'{e}}moire sur les
  lois du mouvement des fluides}},\ }\href@noop {} {\bibfield  {journal}
  {\bibinfo  {journal} {Mem. Acad. Sci. Inst. Fr}\ }\textbf {\bibinfo {volume}
  {6}},\ \bibinfo {pages} {389} (\bibinfo {year} {1823})}\BibitemShut {NoStop}%
\bibitem [{\citenamefont {Neto}\ \emph {et~al.}(2005)\citenamefont {Neto},
  \citenamefont {Evans}, \citenamefont {Bonaccurso}, \citenamefont {Butt},\
  and\ \citenamefont {Craig}}]{Neto2005}%
  \BibitemOpen
  \bibfield  {author} {\bibinfo {author} {\bibfnamefont {C.}~\bibnamefont
  {Neto}}, \bibinfo {author} {\bibfnamefont {D.~R.}\ \bibnamefont {Evans}},
  \bibinfo {author} {\bibfnamefont {E.}~\bibnamefont {Bonaccurso}}, \bibinfo
  {author} {\bibfnamefont {H.-J.}\ \bibnamefont {Butt}},\ and\ \bibinfo
  {author} {\bibfnamefont {V.~S.~J.}\ \bibnamefont {Craig}},\ }\bibfield
  {title} {\bibinfo {title} {{Boundary slip in Newtonian liquids: a review of
  experimental studies}},\ }\href {https://doi.org/10.1088/0034-4885/68/12/R05}
  {\bibfield  {journal} {\bibinfo  {journal} {Rep. Prog. Phys.}\ }\textbf
  {\bibinfo {volume} {68}},\ \bibinfo {pages} {2859} (\bibinfo {year}
  {2005})}\BibitemShut {NoStop}%
\bibitem [{\citenamefont {Maali}\ and\ \citenamefont
  {Bhushan}(2012)}]{Maali2012}%
  \BibitemOpen
  \bibfield  {author} {\bibinfo {author} {\bibfnamefont {A.}~\bibnamefont
  {Maali}}\ and\ \bibinfo {author} {\bibfnamefont {B.}~\bibnamefont
  {Bhushan}},\ }\bibfield  {title} {\bibinfo {title} {{Review article:
  Measurement of slip length on superhydrophobic surfaces}},\ }\href
  {https://doi.org/10.1098/rsta.2011.0505} {\bibfield  {journal} {\bibinfo
  {journal} {Philos. Trans. Royal Soc. A .}\ }\textbf {\bibinfo {volume}
  {370}},\ \bibinfo {pages} {2304} (\bibinfo {year} {2012})}\BibitemShut
  {NoStop}%
\bibitem [{\citenamefont {Lei}\ \emph {et~al.}(2016)\citenamefont {Lei},
  \citenamefont {Rigozzi},\ and\ \citenamefont {McKenzie}}]{Lei2016}%
  \BibitemOpen
  \bibfield  {author} {\bibinfo {author} {\bibfnamefont {W.}~\bibnamefont
  {Lei}}, \bibinfo {author} {\bibfnamefont {M.~K.}\ \bibnamefont {Rigozzi}},\
  and\ \bibinfo {author} {\bibfnamefont {D.~R.}\ \bibnamefont {McKenzie}},\
  }\bibfield  {title} {\bibinfo {title} {{The physics of confined flow and its
  application to water leaks, water permeation and water nanoflows: a
  review}},\ }\href {https://doi.org/10.1088/0034-4885/79/2/025901} {\bibfield
  {journal} {\bibinfo  {journal} {Rep. Prog. Phys.}\ }\textbf {\bibinfo
  {volume} {79}},\ \bibinfo {pages} {025901} (\bibinfo {year}
  {2016})}\BibitemShut {NoStop}%
\bibitem [{\citenamefont {Bocquet}\ and\ \citenamefont
  {Barrat}(1993)}]{Bocquet1993}%
  \BibitemOpen
  \bibfield  {author} {\bibinfo {author} {\bibfnamefont {L.}~\bibnamefont
  {Bocquet}}\ and\ \bibinfo {author} {\bibfnamefont {J.~L.}\ \bibnamefont
  {Barrat}},\ }\bibfield  {title} {\bibinfo {title} {{Hydrodynamic boundary
  conditions and correlation functions of confined fluids}},\ }\href
  {https://doi.org/10.1103/PhysRevLett.70.2726} {\bibfield  {journal} {\bibinfo
   {journal} {Phys. Rev. Lett.}\ }\textbf {\bibinfo {volume} {70}},\ \bibinfo
  {pages} {2726} (\bibinfo {year} {1993})}\BibitemShut {NoStop}%
\bibitem [{\citenamefont {Bocquet}\ and\ \citenamefont
  {Barrat}(1994)}]{Bocquet1994}%
  \BibitemOpen
  \bibfield  {author} {\bibinfo {author} {\bibfnamefont {L.}~\bibnamefont
  {Bocquet}}\ and\ \bibinfo {author} {\bibfnamefont {J.~L.}\ \bibnamefont
  {Barrat}},\ }\bibfield  {title} {\bibinfo {title} {{Hydrodynamic boundary
  conditions, correlation functions, and Kubo relations for confined fluids}},\
  }\href {https://doi.org/10.1103/PhysRevE.49.3079} {\bibfield  {journal}
  {\bibinfo  {journal} {Phys. Rev. E}\ }\textbf {\bibinfo {volume} {49}},\
  \bibinfo {pages} {3079} (\bibinfo {year} {1994})}\BibitemShut {NoStop}%
\bibitem [{\citenamefont {Petravic}\ and\ \citenamefont
  {Harrowell}(2007)}]{Petravic2007}%
  \BibitemOpen
  \bibfield  {author} {\bibinfo {author} {\bibfnamefont {J.}~\bibnamefont
  {Petravic}}\ and\ \bibinfo {author} {\bibfnamefont {P.}~\bibnamefont
  {Harrowell}},\ }\bibfield  {title} {\bibinfo {title} {{On the equilibrium
  calculation of the friction coefficient for liquid slip against a wall}},\
  }\href {https://doi.org/10.1063/1.2799186} {\bibfield  {journal} {\bibinfo
  {journal} {J. Chem. Phys.}\ }\textbf {\bibinfo {volume} {127}},\ \bibinfo
  {pages} {174706} (\bibinfo {year} {2007})}\BibitemShut {NoStop}%
\bibitem [{\citenamefont {Sokhan}\ and\ \citenamefont
  {Quirke}(2008)}]{Sokhan2008}%
  \BibitemOpen
  \bibfield  {author} {\bibinfo {author} {\bibfnamefont {V.~P.}\ \bibnamefont
  {Sokhan}}\ and\ \bibinfo {author} {\bibfnamefont {N.}~\bibnamefont
  {Quirke}},\ }\bibfield  {title} {\bibinfo {title} {{Slip coefficient in
  nanoscale pore flow}},\ }\href {https://doi.org/10.1103/PhysRevE.78.015301}
  {\bibfield  {journal} {\bibinfo  {journal} {Phys. Rev. E}\ }\textbf {\bibinfo
  {volume} {78}},\ \bibinfo {pages} {015301} (\bibinfo {year}
  {2008})}\BibitemShut {NoStop}%
\bibitem [{\citenamefont {Hansen}\ \emph {et~al.}(2011)\citenamefont {Hansen},
  \citenamefont {Todd},\ and\ \citenamefont {Daivis}}]{Hansen2011}%
  \BibitemOpen
  \bibfield  {author} {\bibinfo {author} {\bibfnamefont {J.~S.}\ \bibnamefont
  {Hansen}}, \bibinfo {author} {\bibfnamefont {B.~D.}\ \bibnamefont {Todd}},\
  and\ \bibinfo {author} {\bibfnamefont {P.~J.}\ \bibnamefont {Daivis}},\
  }\bibfield  {title} {\bibinfo {title} {{Prediction of fluid velocity slip at
  solid surfaces}},\ }\href {https://doi.org/10.1103/PhysRevE.84.016313}
  {\bibfield  {journal} {\bibinfo  {journal} {Phys. Rev. E}\ }\textbf {\bibinfo
  {volume} {84}},\ \bibinfo {pages} {016313} (\bibinfo {year}
  {2011})}\BibitemShut {NoStop}%
\bibitem [{\citenamefont {Huang}\ and\ \citenamefont
  {Szlufarska}(2014)}]{Huang2014}%
  \BibitemOpen
  \bibfield  {author} {\bibinfo {author} {\bibfnamefont {K.}~\bibnamefont
  {Huang}}\ and\ \bibinfo {author} {\bibfnamefont {I.}~\bibnamefont
  {Szlufarska}},\ }\bibfield  {title} {\bibinfo {title} {{Green-Kubo relation
  for friction at liquid-solid interfaces}},\ }\href
  {https://doi.org/10.1103/PhysRevE.89.032119} {\bibfield  {journal} {\bibinfo
  {journal} {Phys. Rev. E}\ }\textbf {\bibinfo {volume} {89}},\ \bibinfo
  {pages} {032119} (\bibinfo {year} {2014})}\BibitemShut {NoStop}%
\bibitem [{\citenamefont {Oga}\ \emph {et~al.}(2019)\citenamefont {Oga},
  \citenamefont {Yamaguchi}, \citenamefont {Omori}, \citenamefont {Merabia},\
  and\ \citenamefont {Joly}}]{Oga2019}%
  \BibitemOpen
  \bibfield  {author} {\bibinfo {author} {\bibfnamefont {H.}~\bibnamefont
  {Oga}}, \bibinfo {author} {\bibfnamefont {Y.}~\bibnamefont {Yamaguchi}},
  \bibinfo {author} {\bibfnamefont {T.}~\bibnamefont {Omori}}, \bibinfo
  {author} {\bibfnamefont {S.}~\bibnamefont {Merabia}},\ and\ \bibinfo {author}
  {\bibfnamefont {L.}~\bibnamefont {Joly}},\ }\bibfield  {title} {\bibinfo
  {title} {{Green-Kubo measurement of liquid-solid friction in finite-size
  systems}},\ }\href {https://doi.org/10.1063/1.5104335} {\bibfield  {journal}
  {\bibinfo  {journal} {J. Chem. Phys.}\ }\textbf {\bibinfo {volume} {151}},\
  \bibinfo {pages} {054502} (\bibinfo {year} {2019})}\BibitemShut {NoStop}%
\bibitem [{\citenamefont {Espa{\~{n}}ol}\ \emph {et~al.}(2019)\citenamefont
  {Espa{\~{n}}ol}, \citenamefont {de~la Torre},\ and\ \citenamefont
  {Duque-Zumajo}}]{Espanol2019}%
  \BibitemOpen
  \bibfield  {author} {\bibinfo {author} {\bibfnamefont {P.}~\bibnamefont
  {Espa{\~{n}}ol}}, \bibinfo {author} {\bibfnamefont {J.~A.}\ \bibnamefont
  {de~la Torre}},\ and\ \bibinfo {author} {\bibfnamefont {D.}~\bibnamefont
  {Duque-Zumajo}},\ }\bibfield  {title} {\bibinfo {title} {{Solution to the
  plateau problem in the Green-Kubo formula}},\ }\href
  {https://doi.org/10.1103/PhysRevE.99.022126} {\bibfield  {journal} {\bibinfo
  {journal} {Phys. Rev. E}\ }\textbf {\bibinfo {volume} {99}},\ \bibinfo
  {pages} {022126} (\bibinfo {year} {2019})}\BibitemShut {NoStop}%
\bibitem [{\citenamefont {{De La Torre}}\ \emph {et~al.}(2019)\citenamefont
  {{De La Torre}}, \citenamefont {Duque-Zumajo}, \citenamefont {Camargo},\ and\
  \citenamefont {Espa{\~{n}}ol}}]{DeLaTorre2019}%
  \BibitemOpen
  \bibfield  {author} {\bibinfo {author} {\bibfnamefont {J.~A.}\ \bibnamefont
  {{De La Torre}}}, \bibinfo {author} {\bibfnamefont {D.}~\bibnamefont
  {Duque-Zumajo}}, \bibinfo {author} {\bibfnamefont {D.}~\bibnamefont
  {Camargo}},\ and\ \bibinfo {author} {\bibfnamefont {P.}~\bibnamefont
  {Espa{\~{n}}ol}},\ }\bibfield  {title} {\bibinfo {title} {{Microscopic Slip
  Boundary Conditions in Unsteady Fluid Flows}},\ }\href
  {https://doi.org/10.1103/PhysRevLett.123.264501} {\bibfield  {journal}
  {\bibinfo  {journal} {Phys. Rev. Lett.}\ }\textbf {\bibinfo {volume} {123}},\
  \bibinfo {pages} {264501} (\bibinfo {year} {2019})}\BibitemShut {NoStop}%
\bibitem [{\citenamefont {Nakano}\ and\ \citenamefont
  {Sasa}(2019)}]{Nakano2019a}%
  \BibitemOpen
  \bibfield  {author} {\bibinfo {author} {\bibfnamefont {H.}~\bibnamefont
  {Nakano}}\ and\ \bibinfo {author} {\bibfnamefont {S.-i.}\ \bibnamefont
  {Sasa}},\ }\bibfield  {title} {\bibinfo {title} {{Microscopic determination
  of macroscopic boundary conditions in Newtonian liquids}},\ }\href
  {https://doi.org/10.1103/PhysRevE.99.013106} {\bibfield  {journal} {\bibinfo
  {journal} {Phys. Rev. E}\ }\textbf {\bibinfo {volume} {99}},\ \bibinfo
  {pages} {013106} (\bibinfo {year} {2019})}\BibitemShut {NoStop}%
\bibitem [{\citenamefont {Nakano}\ and\ \citenamefont
  {Sasa}(2020)}]{Nakano2019c}%
  \BibitemOpen
  \bibfield  {author} {\bibinfo {author} {\bibfnamefont {H.}~\bibnamefont
  {Nakano}}\ and\ \bibinfo {author} {\bibfnamefont {S.-i.}\ \bibnamefont
  {Sasa}},\ }\bibfield  {title} {\bibinfo {title} {{Equilibrium measurement
  method of slip length based on fluctuating hydrodynamics}},\ }\href
  {http://arxiv.org/abs/1910.03825
  https://link.aps.org/doi/10.1103/PhysRevE.101.033109} {\bibfield  {journal}
  {\bibinfo  {journal} {Phys. Rev. E}\ }\textbf {\bibinfo {volume} {101}},\
  \bibinfo {pages} {033109} (\bibinfo {year} {2020})}\BibitemShut {NoStop}%
\bibitem [{\citenamefont {Omori}\ \emph {et~al.}(2019)\citenamefont {Omori},
  \citenamefont {Inoue}, \citenamefont {Joly}, \citenamefont {Merabia},\ and\
  \citenamefont {Yamaguchi}}]{Omori2019a}%
  \BibitemOpen
  \bibfield  {author} {\bibinfo {author} {\bibfnamefont {T.}~\bibnamefont
  {Omori}}, \bibinfo {author} {\bibfnamefont {N.}~\bibnamefont {Inoue}},
  \bibinfo {author} {\bibfnamefont {L.}~\bibnamefont {Joly}}, \bibinfo {author}
  {\bibfnamefont {S.}~\bibnamefont {Merabia}},\ and\ \bibinfo {author}
  {\bibfnamefont {Y.}~\bibnamefont {Yamaguchi}},\ }\bibfield  {title} {\bibinfo
  {title} {{Full characterization of the hydrodynamic boundary condition at the
  atomic scale using an oscillating channel: Identification of the viscoelastic
  interfacial friction and the hydrodynamic boundary position}},\ }\href
  {https://doi.org/10.1103/PhysRevFluids.4.114201} {\bibfield  {journal}
  {\bibinfo  {journal} {Phys. Rev. Fluids}\ }\textbf {\bibinfo {volume} {4}},\
  \bibinfo {pages} {114201} (\bibinfo {year} {2019})}\BibitemShut {NoStop}%
\bibitem [{\citenamefont {Slie}\ \emph {et~al.}(1966)\citenamefont {Slie},
  \citenamefont {Donfor~Jr},\ and\ \citenamefont {Litovitz}}]{slie1966}%
  \BibitemOpen
  \bibfield  {author} {\bibinfo {author} {\bibfnamefont {W.}~\bibnamefont
  {Slie}}, \bibinfo {author} {\bibfnamefont {A.}~\bibnamefont {Donfor~Jr}},\
  and\ \bibinfo {author} {\bibfnamefont {T.}~\bibnamefont {Litovitz}},\
  }\bibfield  {title} {\bibinfo {title} {Ultrasonic shear and longitudinal
  measurements in aqueous glycerol},\ }\href@noop {} {\bibfield  {journal}
  {\bibinfo  {journal} {J. Chem. Phys.}\ }\textbf {\bibinfo {volume} {44}},\
  \bibinfo {pages} {3712} (\bibinfo {year} {1966})}\BibitemShut {NoStop}%
\bibitem [{\citenamefont {Masciovecchio}\ \emph {et~al.}(2004)\citenamefont
  {Masciovecchio}, \citenamefont {Santucci}, \citenamefont {Gessini},
  \citenamefont {Di~Fonzo}, \citenamefont {Ruocco},\ and\ \citenamefont
  {Sette}}]{masciovecchio2004}%
  \BibitemOpen
  \bibfield  {author} {\bibinfo {author} {\bibfnamefont {C.}~\bibnamefont
  {Masciovecchio}}, \bibinfo {author} {\bibfnamefont {S.}~\bibnamefont
  {Santucci}}, \bibinfo {author} {\bibfnamefont {A.}~\bibnamefont {Gessini}},
  \bibinfo {author} {\bibfnamefont {S.}~\bibnamefont {Di~Fonzo}}, \bibinfo
  {author} {\bibfnamefont {G.}~\bibnamefont {Ruocco}},\ and\ \bibinfo {author}
  {\bibfnamefont {F.}~\bibnamefont {Sette}},\ }\bibfield  {title} {\bibinfo
  {title} {Structural relaxation in liquid water by inelastic {UV}
  scattering},\ }\href@noop {} {\bibfield  {journal} {\bibinfo  {journal}
  {Phys. Rev. Lett.}\ }\textbf {\bibinfo {volume} {92}},\ \bibinfo {pages}
  {255507} (\bibinfo {year} {2004})}\BibitemShut {NoStop}%
\bibitem [{\citenamefont {Omelyan}\ \emph {et~al.}(2005)\citenamefont
  {Omelyan}, \citenamefont {Mryglod},\ and\ \citenamefont
  {Tokarchuk}}]{Omelyan2005}%
  \BibitemOpen
  \bibfield  {author} {\bibinfo {author} {\bibfnamefont {I.}~\bibnamefont
  {Omelyan}}, \bibinfo {author} {\bibfnamefont {I.}~\bibnamefont {Mryglod}},\
  and\ \bibinfo {author} {\bibfnamefont {M.}~\bibnamefont {Tokarchuk}},\
  }\bibfield  {title} {\bibinfo {title} {{Wavevector- and frequency-dependent
  shear viscosity of water: the modified collective mode approach and molecular
  dynamics calculations}},\ }\href {https://doi.org/10.5488/CMP.8.1.25}
  {\bibfield  {journal} {\bibinfo  {journal} {Condens. Matter Phys.}\ }\textbf
  {\bibinfo {volume} {8}},\ \bibinfo {pages} {25} (\bibinfo {year}
  {2005})}\BibitemShut {NoStop}%
\bibitem [{\citenamefont {O'Sullivan}\ \emph {et~al.}(2019)\citenamefont
  {O'Sullivan}, \citenamefont {Kannam}, \citenamefont {Chakraborty},
  \citenamefont {Todd},\ and\ \citenamefont {Sader}}]{osullivan2019}%
  \BibitemOpen
  \bibfield  {author} {\bibinfo {author} {\bibfnamefont {T.~J.}\ \bibnamefont
  {O'Sullivan}}, \bibinfo {author} {\bibfnamefont {S.~K.}\ \bibnamefont
  {Kannam}}, \bibinfo {author} {\bibfnamefont {D.}~\bibnamefont {Chakraborty}},
  \bibinfo {author} {\bibfnamefont {B.~D.}\ \bibnamefont {Todd}},\ and\
  \bibinfo {author} {\bibfnamefont {J.~E.}\ \bibnamefont {Sader}},\ }\bibfield
  {title} {\bibinfo {title} {Viscoelasticity of liquid water investigated using
  molecular dynamics simulations},\ }\href@noop {} {\bibfield  {journal}
  {\bibinfo  {journal} {Phys. Rev. Fluids}\ }\textbf {\bibinfo {volume} {4}},\
  \bibinfo {pages} {123302} (\bibinfo {year} {2019})}\BibitemShut {NoStop}%
\bibitem [{\citenamefont {Straube}\ \emph {et~al.}(2020)\citenamefont
  {Straube}, \citenamefont {Kowalik}, \citenamefont {Netz},\ and\ \citenamefont
  {H{\"{o}}fling}}]{Straube2020}%
  \BibitemOpen
  \bibfield  {author} {\bibinfo {author} {\bibfnamefont {A.~V.}\ \bibnamefont
  {Straube}}, \bibinfo {author} {\bibfnamefont {B.~G.}\ \bibnamefont
  {Kowalik}}, \bibinfo {author} {\bibfnamefont {R.~R.}\ \bibnamefont {Netz}},\
  and\ \bibinfo {author} {\bibfnamefont {F.}~\bibnamefont {H{\"{o}}fling}},\
  }\bibfield  {title} {\bibinfo {title} {{Rapid onset of molecular friction in
  liquids bridging between the atomistic and hydrodynamic pictures}},\ }\href
  {https://doi.org/10.1038/s42005-020-0389-0} {\bibfield  {journal} {\bibinfo
  {journal} {Commun. Phys.}\ }\textbf {\bibinfo {volume} {3}},\ \bibinfo
  {pages} {126} (\bibinfo {year} {2020})}\BibitemShut {NoStop}%
\bibitem [{\citenamefont {Schulz}\ \emph {et~al.}(2020)\citenamefont {Schulz},
  \citenamefont {Schlaich}, \citenamefont {Heyden}, \citenamefont {Netz},\ and\
  \citenamefont {Kappler}}]{Schulz2020}%
  \BibitemOpen
  \bibfield  {author} {\bibinfo {author} {\bibfnamefont {J.~C.~F.}\
  \bibnamefont {Schulz}}, \bibinfo {author} {\bibfnamefont {A.}~\bibnamefont
  {Schlaich}}, \bibinfo {author} {\bibfnamefont {M.}~\bibnamefont {Heyden}},
  \bibinfo {author} {\bibfnamefont {R.~R.}\ \bibnamefont {Netz}},\ and\
  \bibinfo {author} {\bibfnamefont {J.}~\bibnamefont {Kappler}},\ }\bibfield
  {title} {\bibinfo {title} {{Molecular interpretation of the non-Newtonian
  viscoelastic behavior of liquid water at high frequencies}},\ }\href
  {https://doi.org/10.1103/physrevfluids.5.103301} {\bibfield  {journal}
  {\bibinfo  {journal} {Phys. Rev. Fluids}\ }\textbf {\bibinfo {volume} {5}},\
  \bibinfo {pages} {1} (\bibinfo {year} {2020})},\ \Eprint
  {https://arxiv.org/abs/2003.08309} {2003.08309} \BibitemShut {NoStop}%
\bibitem [{\citenamefont {Bocquet}\ and\ \citenamefont
  {Barrat}(2013)}]{Bocquet2013}%
  \BibitemOpen
  \bibfield  {author} {\bibinfo {author} {\bibfnamefont {L.}~\bibnamefont
  {Bocquet}}\ and\ \bibinfo {author} {\bibfnamefont {J.-L.}\ \bibnamefont
  {Barrat}},\ }\bibfield  {title} {\bibinfo {title} {{On the Green-Kubo
  relationship for the liquid-solid friction coefficient}},\ }\href
  {https://doi.org/10.1063/1.4816006} {\bibfield  {journal} {\bibinfo
  {journal} {J. Chem. Phys.}\ }\textbf {\bibinfo {volume} {139}},\ \bibinfo
  {pages} {044704} (\bibinfo {year} {2013})}\BibitemShut {NoStop}%
\bibitem [{Note1()}]{Note1}%
  \BibitemOpen
  \bibinfo {note} {As we show in the supplemental material, the frequency
  response characteristics of the FC and of the bulk viscosity are similar.
  This means that there is no bulk flow in the time scales where the
  high-frequency mode of the FC such as the elasticity dominates \cite
  {Omori2019a}, and therefore one can neglect the frequency-dependent component
  of the bulk viscosity in the present analysis.}\BibitemShut {Stop}%
\bibitem [{\citenamefont {Mazur}\ and\ \citenamefont
  {Oppenheim}(1970)}]{Mazur1970}%
  \BibitemOpen
  \bibfield  {author} {\bibinfo {author} {\bibfnamefont {P.}~\bibnamefont
  {Mazur}}\ and\ \bibinfo {author} {\bibfnamefont {I.}~\bibnamefont
  {Oppenheim}},\ }\bibfield  {title} {\bibinfo {title} {{Molecular theory of
  Brownian motion}},\ }\href {https://doi.org/10.1016/0031-8914(70)90005-4}
  {\bibfield  {journal} {\bibinfo  {journal} {Physica}\ }\textbf {\bibinfo
  {volume} {50}},\ \bibinfo {pages} {241} (\bibinfo {year} {1970})}\BibitemShut
  {NoStop}%
\bibitem [{\citenamefont {Nishida}\ \emph {et~al.}(2014)\citenamefont
  {Nishida}, \citenamefont {Surblys}, \citenamefont {Yamaguchi}, \citenamefont
  {Kuroda}, \citenamefont {Kagawa}, \citenamefont {Nakajima},\ and\
  \citenamefont {Fujimura}}]{Nishida2014}%
  \BibitemOpen
  \bibfield  {author} {\bibinfo {author} {\bibfnamefont {S.}~\bibnamefont
  {Nishida}}, \bibinfo {author} {\bibfnamefont {D.}~\bibnamefont {Surblys}},
  \bibinfo {author} {\bibfnamefont {Y.}~\bibnamefont {Yamaguchi}}, \bibinfo
  {author} {\bibfnamefont {K.}~\bibnamefont {Kuroda}}, \bibinfo {author}
  {\bibfnamefont {M.}~\bibnamefont {Kagawa}}, \bibinfo {author} {\bibfnamefont
  {T.}~\bibnamefont {Nakajima}},\ and\ \bibinfo {author} {\bibfnamefont
  {H.}~\bibnamefont {Fujimura}},\ }\bibfield  {title} {\bibinfo {title}
  {Molecular dynamics analysis of multiphase interfaces based on in situ
  extraction of the pressure distribution of a liquid droplet on a solid
  surface},\ }\href {https://doi.org/http://dx.doi.org/10.1063/1.4865254}
  {\bibfield  {journal} {\bibinfo  {journal} {J. Chem. Phys.}\ }\textbf
  {\bibinfo {volume} {140}},\ \bibinfo {pages} {074707} (\bibinfo {year}
  {2014})}\BibitemShut {NoStop}%
\bibitem [{\citenamefont {Ogawa}\ \emph {et~al.}(2019)\citenamefont {Ogawa},
  \citenamefont {Oga}, \citenamefont {Kusudo}, \citenamefont {Yamaguchi},
  \citenamefont {Omori}, \citenamefont {Merabia},\ and\ \citenamefont
  {Joly}}]{Ogawa2019}%
  \BibitemOpen
  \bibfield  {author} {\bibinfo {author} {\bibfnamefont {K.}~\bibnamefont
  {Ogawa}}, \bibinfo {author} {\bibfnamefont {H.}~\bibnamefont {Oga}}, \bibinfo
  {author} {\bibfnamefont {H.}~\bibnamefont {Kusudo}}, \bibinfo {author}
  {\bibfnamefont {Y.}~\bibnamefont {Yamaguchi}}, \bibinfo {author}
  {\bibfnamefont {T.}~\bibnamefont {Omori}}, \bibinfo {author} {\bibfnamefont
  {S.}~\bibnamefont {Merabia}},\ and\ \bibinfo {author} {\bibfnamefont
  {L.}~\bibnamefont {Joly}},\ }\bibfield  {title} {\bibinfo {title} {{Large
  effect of lateral box size in molecular dynamics simulations of liquid-solid
  friction}},\ }\href {https://doi.org/10.1103/PhysRevE.100.023101} {\bibfield
  {journal} {\bibinfo  {journal} {Phys. Rev. E}\ }\textbf {\bibinfo {volume}
  {100}},\ \bibinfo {pages} {023101} (\bibinfo {year} {2019})}\BibitemShut
  {NoStop}%
\bibitem [{\citenamefont {Abascal}\ \emph {et~al.}(2005)\citenamefont
  {Abascal}, \citenamefont {Sanz}, \citenamefont
  {Garc{\'{i}}a~Fern{\'{a}}ndez},\ and\ \citenamefont {Vega}}]{abascal2005}%
  \BibitemOpen
  \bibfield  {author} {\bibinfo {author} {\bibfnamefont {J.~L.~F.}\
  \bibnamefont {Abascal}}, \bibinfo {author} {\bibfnamefont {E.}~\bibnamefont
  {Sanz}}, \bibinfo {author} {\bibfnamefont {R.}~\bibnamefont
  {Garc{\'{i}}a~Fern{\'{a}}ndez}},\ and\ \bibinfo {author} {\bibfnamefont
  {C.}~\bibnamefont {Vega}},\ }\bibfield  {title} {\bibinfo {title} {{A
  potential model for the study of ices and amorphous water: TIP4P/Ice.}},\
  }\href {https://doi.org/10.1063/1.1931662} {\bibfield  {journal} {\bibinfo
  {journal} {J. Chem. Phys.}\ }\textbf {\bibinfo {volume} {122}},\ \bibinfo
  {pages} {234511} (\bibinfo {year} {2005})}\BibitemShut {NoStop}%
\bibitem [{\citenamefont {Huang}\ \emph {et~al.}(2008)\citenamefont {Huang},
  \citenamefont {Cottin-Bizonne}, \citenamefont {Ybert},\ and\ \citenamefont
  {Bocquet}}]{Huang2008}%
  \BibitemOpen
  \bibfield  {author} {\bibinfo {author} {\bibfnamefont {D.~M.}\ \bibnamefont
  {Huang}}, \bibinfo {author} {\bibfnamefont {C.}~\bibnamefont
  {Cottin-Bizonne}}, \bibinfo {author} {\bibfnamefont {C.}~\bibnamefont
  {Ybert}},\ and\ \bibinfo {author} {\bibfnamefont {L.}~\bibnamefont
  {Bocquet}},\ }\bibfield  {title} {\bibinfo {title} {{Aqueous Electrolytes
  near Hydrophobic Surfaces: Dynamic Effects of Ion Specificity and
  Hydrodynamic Slip}},\ }\href {https://doi.org/10.1021/la7021787} {\bibfield
  {journal} {\bibinfo  {journal} {Langmuir}\ }\textbf {\bibinfo {volume}
  {24}},\ \bibinfo {pages} {1442} (\bibinfo {year} {2008})}\BibitemShut
  {NoStop}%
\bibitem [{\citenamefont {Falk}\ \emph {et~al.}(2010)\citenamefont {Falk},
  \citenamefont {Sedlmeier}, \citenamefont {Joly}, \citenamefont {Netz},\ and\
  \citenamefont {Bocquet}}]{falk2010}%
  \BibitemOpen
  \bibfield  {author} {\bibinfo {author} {\bibfnamefont {K.}~\bibnamefont
  {Falk}}, \bibinfo {author} {\bibfnamefont {F.}~\bibnamefont {Sedlmeier}},
  \bibinfo {author} {\bibfnamefont {L.}~\bibnamefont {Joly}}, \bibinfo {author}
  {\bibfnamefont {R.~R.}\ \bibnamefont {Netz}},\ and\ \bibinfo {author}
  {\bibfnamefont {L.}~\bibnamefont {Bocquet}},\ }\bibfield  {title} {\bibinfo
  {title} {{Molecular origin of fast water transport in carbon nanotube
  membranes: superlubricity versus curvature dependent friction.}},\ }\href
  {https://doi.org/10.1021/nl1021046} {\bibfield  {journal} {\bibinfo
  {journal} {Nano Lett.}\ }\textbf {\bibinfo {volume} {10}},\ \bibinfo {pages}
  {4067} (\bibinfo {year} {2010})}\BibitemShut {NoStop}%
\bibitem [{\citenamefont {Herrero}\ \emph {et~al.}(2020)\citenamefont
  {Herrero}, \citenamefont {Tocci}, \citenamefont {Merabia},\ and\
  \citenamefont {Joly}}]{Herrero2020}%
  \BibitemOpen
  \bibfield  {author} {\bibinfo {author} {\bibfnamefont {C.}~\bibnamefont
  {Herrero}}, \bibinfo {author} {\bibfnamefont {G.}~\bibnamefont {Tocci}},
  \bibinfo {author} {\bibfnamefont {S.}~\bibnamefont {Merabia}},\ and\ \bibinfo
  {author} {\bibfnamefont {L.}~\bibnamefont {Joly}},\ }\bibfield  {title}
  {\bibinfo {title} {{Fast increase of nanofluidic slip in supercooled water:
  the key role of dynamics}},\ }\href {https://doi.org/10.1039/D0NR06399A}
  {\bibfield  {journal} {\bibinfo  {journal} {Nanoscale}\ }\textbf {\bibinfo
  {volume} {12}},\ \bibinfo {pages} {20396} (\bibinfo {year}
  {2020})}\BibitemShut {NoStop}%
\bibitem [{Note2()}]{Note2}%
  \BibitemOpen
  \bibinfo {note} {The dimensions of the larger system was identical to
  Ref.~\protect \citenum {Omori2019a}. Therefore, the hydrodynamics height of
  the system $h$ of the larger system was taken from Ref.~\protect \citenum
  {Omori2019a}. For the smaller system, $h$ was calculated considering that the
  hydrodynamic wall position should be identical to that of the larger
  system.}\BibitemShut {Stop}%
\bibitem [{\citenamefont {Joly}\ \emph {et~al.}(2016)\citenamefont {Joly},
  \citenamefont {Tocci}, \citenamefont {Merabia},\ and\ \citenamefont
  {Michaelides}}]{Joly2016}%
  \BibitemOpen
  \bibfield  {author} {\bibinfo {author} {\bibfnamefont {L.}~\bibnamefont
  {Joly}}, \bibinfo {author} {\bibfnamefont {G.}~\bibnamefont {Tocci}},
  \bibinfo {author} {\bibfnamefont {S.}~\bibnamefont {Merabia}},\ and\ \bibinfo
  {author} {\bibfnamefont {A.}~\bibnamefont {Michaelides}},\ }\bibfield
  {title} {\bibinfo {title} {{Strong Coupling between Nanofluidic Transport and
  Interfacial Chemistry: How Defect Reactivity Controls Liquid–Solid Friction
  through Hydrogen Bonding}},\ }\href
  {https://doi.org/10.1021/acs.jpclett.6b00280} {\bibfield  {journal} {\bibinfo
   {journal} {J. Phys. Chem. Lett.}\ }\textbf {\bibinfo {volume} {7}},\
  \bibinfo {pages} {1381} (\bibinfo {year} {2016})}\BibitemShut {NoStop}%
\bibitem [{\citenamefont {Herrero}\ \emph {et~al.}(2019)\citenamefont
  {Herrero}, \citenamefont {Omori}, \citenamefont {Yamaguchi},\ and\
  \citenamefont {Joly}}]{Herrero2019}%
  \BibitemOpen
  \bibfield  {author} {\bibinfo {author} {\bibfnamefont {C.}~\bibnamefont
  {Herrero}}, \bibinfo {author} {\bibfnamefont {T.}~\bibnamefont {Omori}},
  \bibinfo {author} {\bibfnamefont {Y.}~\bibnamefont {Yamaguchi}},\ and\
  \bibinfo {author} {\bibfnamefont {L.}~\bibnamefont {Joly}},\ }\bibfield
  {title} {\bibinfo {title} {{Shear force measurement of the hydrodynamic wall
  position in molecular dynamics}},\ }\href {https://doi.org/10.1063/1.5111966}
  {\bibfield  {journal} {\bibinfo  {journal} {J. Chem. Phys.}\ }\textbf
  {\bibinfo {volume} {151}},\ \bibinfo {pages} {041103} (\bibinfo {year}
  {2019})}\BibitemShut {NoStop}%
\bibitem [{\citenamefont {Kob}\ and\ \citenamefont {Andersen}(1995)}]{kob1995}%
  \BibitemOpen
  \bibfield  {author} {\bibinfo {author} {\bibfnamefont {W.}~\bibnamefont
  {Kob}}\ and\ \bibinfo {author} {\bibfnamefont {H.~C.}\ \bibnamefont
  {Andersen}},\ }\bibfield  {title} {\bibinfo {title} {{Testing mode-coupling
  theory for a supercooled binary Lennard-Jones mixture. II. Intermediate
  scattering function and dynamic susceptibility}},\ }\href
  {https://doi.org/10.1103/PhysRevE.52.4134} {\bibfield  {journal} {\bibinfo
  {journal} {Phys. Rev. E}\ }\textbf {\bibinfo {volume} {52}},\ \bibinfo
  {pages} {4134} (\bibinfo {year} {1995})}\BibitemShut {NoStop}%
\bibitem [{\citenamefont {Gallo}\ \emph {et~al.}(1996)\citenamefont {Gallo},
  \citenamefont {Sciortino}, \citenamefont {Tartaglia},\ and\ \citenamefont
  {Chen}}]{Gallo1996}%
  \BibitemOpen
  \bibfield  {author} {\bibinfo {author} {\bibfnamefont {P.}~\bibnamefont
  {Gallo}}, \bibinfo {author} {\bibfnamefont {F.}~\bibnamefont {Sciortino}},
  \bibinfo {author} {\bibfnamefont {P.}~\bibnamefont {Tartaglia}},\ and\
  \bibinfo {author} {\bibfnamefont {S.~H.}\ \bibnamefont {Chen}},\ }\bibfield
  {title} {\bibinfo {title} {{Slow Dynamics of Water Molecules in Supercooled
  States}},\ }\href {https://doi.org/10.1103/PhysRevLett.76.2730} {\bibfield
  {journal} {\bibinfo  {journal} {Phys. Rev. Lett.}\ }\textbf {\bibinfo
  {volume} {76}},\ \bibinfo {pages} {2730} (\bibinfo {year}
  {1996})}\BibitemShut {NoStop}%
\end{thebibliography}
%

\end{document}